\documentclass[journal]{IEEEtran}
\usepackage{cite}
\usepackage{empheq}
\usepackage{amsmath}
\usepackage{amsthm}
\usepackage{amssymb}
\usepackage{mathrsfs}
\usepackage{ulem}
\usepackage{enumerate}
\usepackage{graphicx}

\graphicspath{{Figs/}{../pdf/}{../jpeg/}}
\DeclareGraphicsExtensions{.pdf,.jpeg,.png,.jpg}

\usepackage{cancel}
\usepackage{makecell}
\usepackage{bbding}
\usepackage{booktabs}
\usepackage{array}
\usepackage{bm}

\usepackage{supertabular}

\newtheorem{remark}[]{Remark}

\usepackage{xcolor}

\definecolor{blue_matlab}{RGB}{0,48,189}
\definecolor{red_matlab}{RGB}{217,83,25}
\definecolor{yellow_matlab}{RGB}{237,177,32}

\newcommand{\tr}[0]{\color{red}}

\begin{document}

\title{
	% Frequency/Voltage Response Analysis and Strength Assessment in CIG-integrated Power Systems: \\ From  the Perspective of Modal Subsystem
	% Frequency and Voltage Strength Analysis of Power Electronics-Dominated Power Systems \\ Based on Modal Decoupling 
	Analysis of Frequency and Voltage Strength in Power Electronics-Dominated Power Systems \\ Based on Eigen-subsystems 

}

\author{
	Huisheng~Gao,
	Linbin~Huang, {\it Member, IEEE},
	Huanhai~Xin, {\it Senior Member, IEEE}, \\
        % Quanmao~Li,
    % Wei~Dong, \\
    % Ying~Yang,
	Zhiyi~Li, {\it Senior Member, IEEE}
	and Ping~Ju, {\it Senior Member, IEEE}
	\vspace{-5mm}
	\thanks{ 
	% Huisheng~Gao, Linbin~Huang, Huanhai~Xin, Zhiyi~Li and Ping~Ju 
	The authors are with the College of Electrical Engineering, Zhejiang University, Hangzhou, China (Emails: \text\{gaohuisheng, hlinbin, xinhh, zhiyi, pju\}@zju.edu.cn). Huanhai~Xin is also with the Zhejiang Key Laboratory of Electrical Technology and System on Renewable Energy, Hangzhou, China.	
	% Wei~Dong and Ying~Yang are with the State Grid Zhejiang Electric Power Research Institute, Hangzhou, China (Emails: eedongwei@163.com, 48458179@qq.com). 

	}

		% Quanmao~Li is with the State Grid Gansu Electric Power Research Institute, Lanzhou, China (Emails: 11369394@qq.com). 
}
\vspace{-5mm}
\maketitle

\begin{abstract}
	The large-scale integration of inverter-based resources (IBRs) has deteriorated the frequency/voltage (F/V) responses of power systems, leading to a higher risk of instability. Consequently, evaluating the F/V strength has become an important task in power electronics (PE)-dominated power systems. Existing methods typically examine F/V strength separately, employing fundamentally different metrics, such as inertia (focusing on device dynamics) and short-circuit ratio (SCR, addressing network characteristics). These fragmented approaches have resulted in a lack of comprehensive understanding of the overall system strength, potentially overlooking critical aspects.
	To address this problem, this paper proposes a unified framework for analyzing F/V strength. First, a unified modeling of F/V regulations is introduced. Then, based on modal decoupling, the power systems are decomposed into several eigen-subsystems, where the F/V responses are both decomposed into common-mode (CM) and differential-mode (DM) components, namely, CM-F, DM-F, CM-V, and DM-V.
	The CM-F and CM-V represent the collective response of all devices to external active or reactive power disturbances, independent of the power network characteristics. In contrast, the DM-F and DM-V capture the redistribution of disturbance power within the system, which is strongly influenced by the network topology and the locations of devices.
	Notably, traditional strength analysis generally ignores the CM-V (global voltage response), which, as discovered in this paper, may also become unstable in PE-dominated power systems.
	Based on the proposed framework, new metrics are proposed to evaluate the strength of each modal component. Finally, the effectiveness of the proposed approach is validated through simulations.

	% These modal strength metrics are compared with traditional metrics, such as total inertia and SCR, to underscore the importance of distinguishing different modes when assessing system strength.
\end{abstract}
	

\section*{Nomenclature}

\subsection{Abbreviations}

\begin{table}[h]\normalsize
\begin{tabular}{ ll } 
	IBR & Inverter-Based Resource \\
	PE & Power Electronic \\
	F/V & Frequency/Voltage \\

	PFR/SFR & Primary/Secondary Frequency Regulation  \\

	SCR & Short-Circuit Ratio \\
	MISCR & Multi-Infeed SCR \\
	MRSCR & Multi-Renewable energy station SCR \\
	gSCR  & generalized SCR \\

	CM/DM  &  Common/Differential Mode \\
	CM-F/CM-V & CM-Frequency/Voltage \\
	DM-F/DM-V & DM-Frequency/Voltage \\

	RoCoF/RoCoV & Rate of Change of Frequency/Voltage \\

	VSG & Virtual Synchronous Generator \\

\end{tabular}
\end{table}

\subsection{Notations}
	We denote by $\bf{0}$ and $\bf{1}$ the column vectors whose elements are respectively all 0 and 1, with appropriate dimensions. Given a set of variables $x_i$, ${\rm diag}\{ x_i \}$ denotes the diagonal matrix where the $i$-th diagonal element is $x_i$. We use subscript ``$e$'' to denote the steady-state value of a variable.
	We use “$s$” to denote the Laplace variable. 
	% For notational simplicity, “$s$” is omitted when appropriate. 

\section{Introduction}\label{sec:intro}

	With the large-scale integration of IBRs, such as renewable energy sources, modern power systems are transitioning toward being dominated by PE.
	The IBRs are generally characterized by limited disturbance rejection capabilities. Hence, their large-scale integration will lead to larger F/V fluctuations in power systems when subject to disturbances, that is, the system becomes weaker and weaker, which increases the risk of instability and significantly constrains the further development of renewable energy~\cite{milano2018foundations}.

	With this regard, the concept of power system strength has gained significant attention in recent years \cite{urdal2015systemstrength,gu2019review}. The term ``strength'' has a long-established research history in mechanics, where it refers to the ability of a material or structure to resist damage under external forces \cite{ross1999strength}. 
	In the power systems context, system strength refers to the ability of a power system to reject disturbance and maintain stability. In early studies, system strength was used to assess the AC system’s ability to accommodate HVDC systems~\cite{kundur1994power}. It was further pointed out in \cite{kundur1994power} that the system strength encompasses two aspects: effective inertia and SCR. Although not explicitly stated, these two aspects correspond to frequency strength and voltage strength. Such a classification has been adopted in several recent studies\cite{ieee2022strength,xu2022systemstrength,sun2024systemstrength}. 

	Regarding the frequency strength, the most common metrics are the system's total inertia and its variants. Due to the increasing spatial variations in frequency responses~\cite{wang2024analytical}, nodal or regional inertia has gained attention in recent years, which reflects the system's local frequency strength \cite{bi2024inertia,zhang2022inertia}. Beyond inertia, PFR or droop control is also an important frequency-supporting resource. However, the PFR performance varies significantly among different types of PE devices. To enable a quantitative comparison, Ref.~\cite{gao2022common} proposed a unified structural model, which approximates the complex frequency support dynamics of different devices using three standard parameters: inertia, damping, and spring constant.

	Research works on the voltage strength are usually based on the concept of SCR, such as MISCR~\cite{davies2007MSCR}, MRSCR \cite{sun2021MRSCR}, gSCR \cite{zhou2023hetero}, and so on \cite{wang2025SynCon}. These works extend the traditional SCR from single-infeed systems to multi-infeed systems, with some applications not limited to HVDC systems but also considering renewable energies. Apart from SCR, Ref.~\cite{wang2024voltagestiffness} introduces the concept of voltage stiffness, namely, the ratio of bus voltage magnitude change before and after a device is connected, which can also be linked to the SCR-based analysis.

    It is worth mentioning that, beyond the aforementioned classification of system strength (i.e., frequency strength and voltage strength), some studies consider that system strength pertains solely to voltage, while inertia is regarded as a separate concept parallel to system strength \cite{operator2018system,gu2019review}. This perspective arises from the distinct response characteristics and quantification metrics of frequency and voltage. To be specific, frequency strength (for convenience, it is still referred to as such) metrics, such as inertia, primarily capture the device characteristics. In contrast, voltage strength metrics, typically based on SCR, emphasize the network characteristics. Moreover, frequency strength considers both global and local characteristics, while voltage strength predominantly focuses on local properties. Note that global frequency metrics, like total inertia, are independent of device locations, while voltage strength metrics are all highly sensitive to this information.

\begin{figure}[!t]
    \centering
    \includegraphics[width=0.95\linewidth]{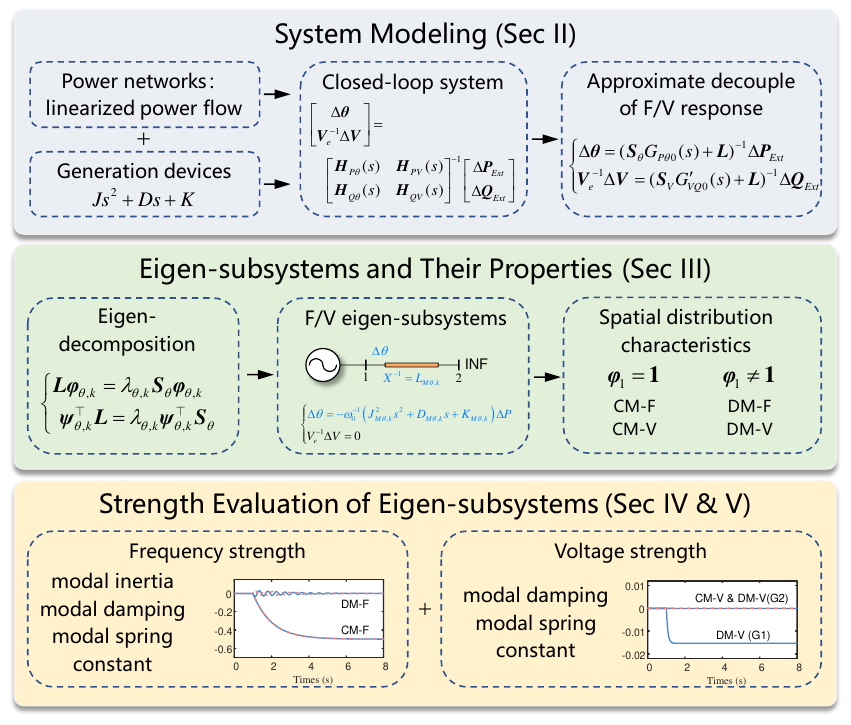}
    \vspace{-3mm}
    \caption{Framework of frequency and voltage strength evaluation.}
    \vspace{-2mm}
    \label{Fig_paper_org}
\end{figure}

	In fact, frequency (phase angle) and voltage (amplitude) compose the voltage phasor. This raises several questions: why are there significant differences in the frequency and voltage responses and their strength metrics? Can these two types of strength be considered within a unified framework?
	To answer these questions, this paper presents a comprehensive analysis of the F/V strength of power systems, as shown in Fig.~\ref{Fig_paper_org}. First, a unified structure (inertia-damper-spring model) is adopted to characterize the F/V regulation dynamics of grid-connected devices. Subsequently, through modal decoupling, the CM and DM components of F/V responses are derived, termed as CM-F, DM-F, CM-V, and DM-V. The CM components represent the global response of the frequency and the voltage, whereas the DM components capture their spatial variations. Based on this framework, modal F/V strength metrics are introduced, elucidating key properties associated with F/V dynamics.
	The main contributions of this paper are as follows:

	1) A unified framework for F/V strength analysis is proposed, theoretically illustrating similarities and differences in F/V responses. A byproduct is the explanation for why frequency responses typically exhibit global characteristics, whereas voltage responses tend to be localized.

	2) The distinct stability mechanisms associated with the CM and DM components of F/V are revealed. Specifically, the stability of CM components depends solely on the total active or reactive power supports of devices across the system, whereas the DM components are heavily influenced by the transmission network and the spatial distribution of devices.

	3) The stability mechanism of the CM-V is investigated for the first time. Traditional voltage strength analysis generally assumes the presence of a strong voltage source (an infinite bus) in the system, which renders the stability problem of CM-V negligible. However, we show that in PE-dominated power systems, IBRs may fail to provide voltage support under disturbances, potentially leading to CM-V instabilities.

	The remainder of the paper is organized as follows: Section~II establishes the closed-loop F/V response model of power systems, where a unified structure is used to simplify the device models. Section~III introduces the F/V eigen-subsystem decoupling method and discusses the global or local characteristics of F/V modal components. Section~IV presents F/V modal strength metrics, and compares them with existing ones. Section~V provides illustrative examples of F/V modal components and the corresponding metrics. Time-domain simulations with high-fidelity models are provided in Section VI. Finally, Section VII presents the conclusions and discusses the assumptions in this paper.

\section{System Modeling}

	This section presents the modeling of F/V responses. The grid-connected devices and the power network are modeled separately and then integrated to form the closed-loop system.

	% Unlike traditional Jacobian matrices that use active/reactive power as interfaces, we adopt normalized active/reactive power~\cite{he2024complex}, which enables the network Jacobian matrix to achieve a symmetric form, facilitating a unified analysis of F/V strength. 
	% This approach enables the network Jacobian matrix to achieve a symmetric form, which will be detailed further below. 

\subsection{Modeling of Power Network}
	Consider an $n$-port network where internal buses are eliminated via Kron reduction \cite{dorfler2012kron}, leaving only terminal (device) buses. 
% 	Let $P_{i}$ and $Q_{i}$ denote the active/reactive power injected into bus $i$, and $V_i$ its voltage amplitude. The normalized active/reactive power ($P_{i}$, $ Q_{i}$) are defined as follows \cite{he2024complex}:
% \begin{equation} \label{Eq_normalized_power}
% 	P_{i} = \dfrac{P_{i}} {V_i^2}, \quad  Q_{i} = \dfrac{Q_{i}} {V_i^2} \,.
% \end{equation}
% 	For simplicity, $P_{i}$ and $ Q_{i}$ will hereinafter be referred to as active and reactive power when no ambiguity arises.
    Linearize the power flow equation and then the network Jacobian matrix can be expressed as:
\begin{equation} \label{Eq_Net_Jacobian}
	\begin{bmatrix} \Delta {\bm P} \\ \Delta {\bm Q} \end{bmatrix}
	=
	\begin{bmatrix} {\bm L} & {\bm N} + 2 {\bm P}_e \\ -{\bm N}  & {\bm L} + 2 {\bm Q}_e \end{bmatrix}
	\begin{bmatrix} \Delta {\bm \theta} \\ {\bm V}_e^{-1} \Delta {\bm V} \end{bmatrix}\,,
\end{equation}
	where $\Delta$ represents the perturbation in the variables, $\Delta {\bm P} $ and $\Delta {\bm Q}$ are column vectors of bus active/reactive power injection $\Delta {P_i}/ \Delta { Q_i}$, $\Delta {\bm \theta}$ and $\Delta {\bm V} $ are column vectors of the phase angles $\Delta {\theta_i}$ and the voltage magnitudes $\Delta {V_i} $. The matrices ${\bm P}_e  = \text{diag} \{  P_{e,i} \}$, ${\bm Q}_e  = \text{diag} \{  Q_{e,i} \}$, ${\bm V}_e  = \text{diag} \{  V_{e,i} \}$, denoting the steady-state acitve power, reactive power, and voltage amplitude at the buses, respectively. The matrices ${\bm L}$ and ${\bm N}$ are expressed as follows \cite{xin2016gSCR_PED,gao2024vcmf}
\begin{equation} \label{Eq_Net_Jacobian_Stat}
	\left\{
	\begin{array}{l}
		L_{ii} =   \sum\nolimits_{j\in i}  V_{e,j}V_{e,i}  (B_{ij} \cos\theta_{e,ij} - G_{ij} \sin \theta_{e,ij}) \,,
	\vspace{1ex}\\
{}		L_{ij} =   -V_{e,j}V_{e,i}  ( B_{ij} \cos\theta_{e,ij} - G_{ij} \sin \theta_{e,ij}) \,,
	\vspace{1ex}\\
		N_{ii}  =   \sum\nolimits_{j\in i}  {V_{e,j}}{V_{e,i}}  (-B_{ij} \sin\theta_{e,ij} - G_{ij} \cos\theta_{e,ij}) \,,
	\vspace{1ex}\\
		N_{ij} =   -{V_{e,j}}{V_{e,i}}  ( -B_{ij} \sin\theta_{e,ij} - G_{ij} \cos \theta_{e,ij}) \,,
	\end{array}
	\right.
\end{equation}
	where $G_{ij}$ and $B_{ij}$ are the conductance and susceptance between buses $i$ and $j$.
	According to \eqref{Eq_Net_Jacobian_Stat}, the row sums of ${\bm L}$ and ${\bm N}$ are all zero, which means, ${\bm 1}$ is a right eigenvector of both ${\bm L}$ and ${\bm N}$, with an associated eigenvalue of 0, i.e.,
\begin{equation} \label{Eq_invariance}
	{\bm 0} = {\bm L} {\bm 1} = {\bm N} {\bm 1}\,.
\end{equation}
	This property indicates the {\it invariance of power flow}, which means, the power flow remains unchanged when all phase angles rotate together ($\Delta \theta_{i} =  \Delta \theta_{j}$). Notably, if the power matrices ($2{\bm P}_e$ and $2{\bm Q}_e$) are moved to the device side (which will be explained in Section II-D), the network matrix will also have zero eigenvalue w.r.t voltage. In fact, it is similar to the property of normalized power flow \cite{he2024complex}, and it will serve as the foundation for deriving CM-F/CM-V in the next section.

	It is worth noting that when the system is lightly-loaded and the network is inductive (i.e., $\theta_{e,ij} \approx 0$ and $G_{ij} \approx 0$), we have ${\bm N} \approx {\bm N} -2 {\bm P}_e \approx {\bm 0}$.
	Then,  $\Delta {\bm P}$ is influenced only by $\Delta {\bm \theta}$, while $\Delta {\bm  Q}$ depends only on ${\bm V}_e^{-1} \Delta {\bm V}$. 
	It means frequency and voltage are approximately decoupled in the network side.

\begin{remark}
This study employs a static power network model, which is appropriate for analyzing F/V responses in the low-frequency range (e.g., below 10 Hz) \cite{gao2024vcmf}. However, it is worth noting that the analysis methods proposed in this paper remain applicable with dynamics network model if it is predominantly inductive \cite{xin2016gSCR_PED}. It is considered as a future work direction.
\end{remark}

	% the invariance of power flow holds not only in linearized cases, but also in nonlinear situations. The invariance with respect to phase angle is evident, while the invariance with respect to voltage can be referenced in [].

\subsection{Modeling of Generation Devices}

	The Jacobian (transfer function) matrix of the devices can be written as:
\begin{equation} \label{Eq_device_Jacobian}
	\begin{bmatrix} \Delta {\bm P} \\ \Delta {\bm  Q} \end{bmatrix}
	= -
	\begin{bmatrix} {\bm G}_{P \theta}(s) & {\bm G}_{PV}(s)  \\ {\bm G}_{Q\theta}(s) & {\bm G}_{QV}(s) \end{bmatrix}
	\begin{bmatrix} \Delta {\bm \theta} \\ {\bm V}_e^{-1} \Delta {\bm V} \end{bmatrix} \,,
\end{equation}
	where ${\bm G}_{\star}(s) = \text{diag} \{ {G}_{\star,i}(s) \}$ represents the device dynamics  (subscript $\star \in \{ P\theta, PV, Q\theta , QV\}$ denotes the input/output). 
	% The negative sign indicates that the devices typically output positive active/reactive power with a drop in the frequency or voltage.

% \begin{figure}[!t]
%     \centering
%     \includegraphics[width=0.95\linewidth]{Fig_FPQV} %
%     \vspace{-3mm}
%     \caption{Demonstration of F-P and Q-V dynamics and their simplifications.}
%     \vspace{-2mm}
%     \label{Fig_FPQV}
% \end{figure}

	Due to the high-order characteristics of devices (particularly the IBRs), a detailed analysis for large-scale power systems is challenging. To make the analysis tractable, the device dynamics are simplified to capture the direct and substantial impact on the F/V responses, as detailed below. 

	The device dynamics that give rise to ${G}_{P\theta}(s) $ for typical SGs or IBRs are expressed as
\begin{equation} \label{Eq_f_rho_tf}
	{G}_{P\theta}(s) 
		= \dfrac{1} {\omega_0} 
			\left( Js^2 + Ds + \dfrac{ K_Ps + K_S } { T_{G}s + 1 } \right)\,,
\end{equation}
	where $\omega_0$ is the nominal frequency; $J$ and $D$ represent the inertia and damping, respectively; $K_P$ and $K_S$ are the gains for PFR and SFR, and $T_{G}$ is a delay constant. For IBRs, their PFRs are fast and can be approximately modeled with  $T_{G}  \approx 0$.

	For the Q-V loop, common control strategies include constant reactive power control, Q-V droop control, and constant voltage control. They can all be represented as in \eqref{Eq_V_sigma_tf}, where $K_{QV}$ is the Q-V droop coefficient, and $T_M$ is the time constant of the low-pass filter. By setting $K_{QV} = 0$ or $K_{QV} = \infty$, it can represent the constant reactive power control or constant voltage control, respectively. 
\begin{equation} \label{Eq_V_sigma_tf}
	{G}_{QV}(s) = K_{QV}(T_M  s + 1)   \,.
\end{equation}

	% Traditionally, reactive power are regulated by voltage and active power are regulated by frequency in the control design of generation devices. Thus, we assume ${G}_{PV}(s) \approx 0$ and ${G}_{\theta Q}(s) \approx 0$ in the analysis. 

	In practice, the control strategy of a device introduces negligible coupling between frequency and reactive power as well as between voltage and active power. As a result, the approximations ${G}_{PV}(s) \approx  {G}_{Q\theta}(s) \approx {0}$ are often valid.
% \begin{equation} \label{Eq_V_sigma_theta_rho_0}
% 	{G}_{PV}(s) \approx  {G}_{Q\theta}(s) \approx {0}\,.
% \end{equation}
	% We have $2 P_e \approx 0$ when the active power output of one single device is far below the transmission limit of the network.

\subsection{Unified Simplification of Devices}
	
\begin{table}[!t]
		\centering
		\caption{Relationship Between Frequency, Voltage and \\ Displacement, Velocity, Acceleration} 
		\vspace{-2mm}
		\setlength{\tabcolsep}{1mm}
		% \resizebox{\columnwidth}{!}{
		\begin{tabular}{ *{4}{c} } 
		\toprule[0.5mm]
		\midrule
				Displacement & Velocity & Acceleration & External Force \\
		\midrule
				$x $ 	& $v = \dot{x}$ & $a = \dot{v}$ & $y_{Ext}$ \\
				$\Delta\theta$ & $ {\tr \Delta\omega} = \Delta \dot{\theta}/\omega_0$ & $ \Delta \dot{\omega}$ & $\Delta P_{Ext}$ \\
				$\tr \Delta V/V_e$ 	& $\Delta\dot{V}/V_e$ & \textbackslash & $\Delta Q_{Ext}$ \\
		\bottomrule[0.5mm]
		\end{tabular}
		% }
		\label{Tab_relation_xva}
\end{table}

	The F/V control strategies described in \eqref{Eq_f_rho_tf} and \eqref{Eq_V_sigma_tf} can be uniformly approximated by a {\it inertia-damper-spring model}, with the method proposed in \cite{gao2022common}, as follows
\begin{equation} \label{Eq_unified_structure}
\left\{
\begin{array}{l}
	{G}_{P\theta}(s) = \omega_0^{-1} ( J_{P\theta} s^2 + D_{P\theta} s + K_{P\theta} ) \,,
	\vspace{1ex}\\
	{G}_{QV}(s)  =  D_{QV} s + K_{QV} \,,
\end{array}
\right.
\end{equation}
	where $J$, $D$, and $K$ are the inertia, damping and spring constant, respectively, and the subscript denotes the input and output. Some parameters can be directly related to those in \eqref{Eq_f_rho_tf} and \eqref{Eq_V_sigma_tf}, e.g., $J_{P\theta} = J$ and $D_{QV} = T_M K_{QV}$. 
	Using the approach proposed in \cite{gao2022common}, the PFR and SFR of an SG can also be approximated by the unified model. 
	
	To further simplify the derivation, it is assumed that the devices have homogeneous parameters:  
\begin{equation} \label{Eq_uniform_devices}
\left\{
	\begin{array}{l}
		{\bm G}_{P \theta}(s) =  {\bm S}_{\theta} G_{P\theta 0}(s) 
	\vspace{1ex}\\
		\quad = \omega_0^{-1} {\bm S}_{\theta}  ( J_{P\theta 0} s^2 + D_{P\theta 0} s + K_{P\theta 0} ) \,,
	\vspace{1ex}\\
		{\bm G}_{QV}(s) =  {\bm S}_{V} G_{QV 0}(s)
	% \vspace{1ex}\\
		= {\bm S}_{V}  ( D_{QV 0} s + K_{QV 0} ) \,,	
	\end{array}
\right.
\end{equation}
	where ${\bm S}_{\theta} = \text{diag}\{ {S}_{\theta,i} \}$ and ${\bm S}_{V} = \text{diag}\{ {S}_{V,i} \}$ represent the relative F/V support capacities of devices, respectively; $G_{P\theta 0}(s)$, $G_{QV 0}(s)$ denote the nominal F/V support dynamics. 
	% For example, if G1 is chosen as the nominal device, then $G_{P\theta 0}(s) = G_{P\theta,1}(s)$ with $S_{\theta,1} = 1$. And $S_{\theta,i}$ can be determined by the relative frequency support capacity of the $i$-th device w.r.t. G1. 
	Note that ${\bm S}_{\theta}$ is not necessarily identical to ${\bm S}_{V}$, e.g., a device with strong voltage control may provide little frequency support. Meanwhile, due to the homogeneous assumption, the steady-state power matrices in \eqref{Eq_Net_Jacobian} can be expressed as 
\begin{equation} \label{Eq_uniform_power}
	{\bm P}_e = {\bm S}_\theta P_{e}, \quad {\bm Q}_e = {\bm S}_V Q_{e},
\end{equation}
where $P_{e}/Q_{e}$ is the nominal active/reactive power of devices.

	\begin{remark}

	This study focuses on the basic properties of F/V strength, and the unified structure in \eqref{Eq_uniform_devices} provides a basis for such investigation. In fact, it aligns with mass-damper-spring models in mechanics \cite{nayfeh2008structural}. In this analogy,  $\Delta\theta$ and  $\Delta V / V_e$ correspond to displacement $x$; $\Delta\omega$ and RoCoV $\Delta \dot{V} / V_e$ correspond to velocity $v = \dot{x}$; RoCoF $\Delta\dot\omega$ corresponds to acceleration $a = \dot{v}$, and power disturbance $\Delta P_{Ext}$, $ \Delta Q_{Ext}$ correspond to external force $y_{Ext}$. These relationships are summarized in TABLE~\ref{Tab_relation_xva}. In this context, inertia, damping, and spring describe the device's ability to reduce acceleration, velocity, and displacement under disturbances, respectively.  

	\end{remark}

	% A comparison between ${G}_{P\theta}(s)$ and ${G}_{QV}(s)$ reveals that the parameters in ${G}_{P\theta}(s)$ are scaled by $\omega_0^{-1}$ w.r.t. ${G}_{QV}(s)$. This difference arises because frequency corresponds to velocity, with its relationship to displacement (phase angle) involving a factor of $\omega_0 / s$, whereas voltage directly corresponds to displacement. Such parameter settings align with conventional practices in power system analysis.

\subsection{Closed-loop System}

	By combining \eqref{Eq_Net_Jacobian} and \eqref{Eq_device_Jacobian}, and incorporating external power disturbances $\Delta {\bm P}_{Ext}/\Delta {\bm  Q}_{Ext}$, the closed-loop responses can be expressed as follows (refer to \cite{gao2024vcmf} for detailed derivations).
\begin{equation} \label{Eq_closed_loop_response}
	\begin{array}{l}
	 	\begin{bmatrix} \Delta {\bm \theta} \\ {\bm V}_e^{-1} \Delta {\bm V} \end{bmatrix}
		=  	\begin{bmatrix} {\bm H}_{P\theta}(s) & {\bm H}_{PV}(s)  \\ {\bm H}_{Q\theta}(s) & {\bm H}_{QV}(s) \end{bmatrix} ^{-1}
		\begin{bmatrix} \Delta {\bm P}_{Ext} \\ \Delta {\bm  Q}_{Ext} \end{bmatrix}
	\vspace{1ex}\\
		\qquad \qquad \quad \approx \begin{bmatrix} {\bm H}_{P\theta}(s) &   \\  & {\bm H}_{QV}(s) \end{bmatrix} ^{-1}
		\begin{bmatrix} \Delta {\bm P}_{Ext} \\ \Delta {\bm  Q}_{Ext} \end{bmatrix}\,,
	\end{array}
\end{equation}
	where ${\bm H}_{P\theta}(s) = {\bm G}_{P \theta}(s) + {\bm L}$, ${\bm H}_{QV}(s) = {\bm G}_{QV}(s) + 2{\bm Q}_e + {\bm L}$, ${\bm H}_{PV}(s) = {\bm G}_{PV}(s) + 2{\bm P}_e + {\bm N}$, ${\bm H}_{Q\theta}(s) = {\bm G}_{Q\theta}(s) - {\bm N}$. 

	Since ${\bm G}_{PV}(s)$, ${\bm G}_{Q\theta}(s)$ and ${\bm N}$ are generally small, we assume ${\bm H}_{Q\theta}(s) \approx {\bm 0}$ and ${\bm H}_{V P}(s) \approx {\bm 0}$ to simplify the analysis. That is, frequency and voltage are decoupled. This simplification facilitates the understanding of the basic principles governing F/V responses. However, it is important to acknowledge that F-V coupling may be strong in specific scenarios, e.g., network is heavily loaded  or devices are equipped with special control \cite{gao2024vcmf}. We will briefly discuss the approach to deal with such coupling in Section VII.

	Combining \eqref{Eq_uniform_devices}-\eqref{Eq_uniform_power}, the F/V responses \eqref{Eq_closed_loop_response} are rewritten as
\begin{equation} \label{Eq_F_V_response_decoupling}
\left\{
	\begin{array}{l}
		\Delta {\bm \theta} = 
			( {\bm S}_{\theta} G_{P\theta 0}(s) + {\bm L})^{-1}  \Delta {\bm P}_{Ext}\,,
	\vspace{1ex}\\
		{\bm V}_e^{-1} \Delta {\bm V} = 
			( {\bm S}_{V} G'_{VQ  0}(s) + {\bm L})^{-1}  \Delta {\bm  Q}_{Ext}\,.
	\end{array}
\right.
\end{equation}
where $G'_{VQ  0}(s) = G_{VQ  0}(s) + 2 Q_{e}$, that is, the power matrix ${\bm Q}_e$ is transferred from the network side to the device side. This transformation introduces symmetry between frequency and voltage, thereby facilitating a unified analysis.

\section{Eigen-subsystems of F/V Responses}

	Although the system in \eqref{Eq_F_V_response_decoupling} has been simplified, analyzing its F/V responses and associated strength remains challenging due to complex interactions among devices. To address this, this section introduces a method for decoupling the multi-device power system into multiple eigen-subsystems, each with only one device. It facilitates the analysis of F/V responses and offers deeper insights into their characteristics.

\subsection{Decoupling Process}

	Since both frequency and voltage transfer function matrices in \eqref{Eq_F_V_response_decoupling} share the same structure, the decoupling process will be  demonstrated using  frequency as an example, and it also applies to voltage.

	Considering the matrix pencil $({\bm L}, {\bm S}_{\theta})$, let $\lambda_{\theta,k}$, ${\bm \psi}_{\theta,k}$ and ${\bm \varphi}_{\theta,k}$ denote its $k$-th generalized eigenvalue and the left/right eigenvector, respectively, i.e.,
\begin{equation} \label{Eq_generalized_eigenvalue}
	{\bm L} {\bm \varphi}_{\theta,k} = \lambda_{\theta,k}  {\bm S}_{\theta} {\bm \varphi}_{\theta,k}, \quad
	{\bm \psi}_{\theta,k}^\top {\bm L}  = \lambda_{\theta,k}  {\bm \psi}_{\theta,k}^\top {\bm S}_{\theta}\,.
\end{equation}
	Arranging \( {\bm \varphi}_{\theta,k} \) and \( {\bm \psi}_{\theta,k} \) to form  \( {\bm \varPhi}_{\theta} = [{\bm \varphi}_{\theta,1}, \dots, {\bm \varphi}_{\theta,n}] \) and \( {\bm \varPsi}_{\theta} = [{\bm \psi}_{\theta,1}, \dots, {\bm \psi}_{\theta,n}] \), and utilizing the orthogonality properties \cite{nayfeh2008structural}, the following matrices are diagonal:
\begin{equation} \label{Eq_F_modal_matrix}
\left\{ \begin{array} {l}
	{\bm S}_{M\theta} := {\bm \varPsi}_\theta^\top {\bm S}_\theta {\bm \varPhi}_\theta = \text{diag}\{ S_{M\theta,k} \}
	\vspace{1ex}\,,\\
	{\bm L}_{M\theta} := {\bm \varPsi}_\theta^\top {\bm L}  {\bm \varPhi}_\theta = \text{diag}\{L_{M\theta,k} \} \,.
\end{array} \right.
\end{equation}
	% where ${\bm S}_{M\theta}$ and ${\bm L}_{M\theta}$ are the modal frequency regulation capacity matrix and modal , respectively.

	Then, the phase angle response in \eqref{Eq_F_V_response_decoupling} can be decoupled as
\begin{equation} \label{Eq_theta_decouple}
	\begin{array}{l}
		\Delta {\bm \theta} =  {\bm \varPhi}_\theta {\bm \varPhi}^{-1}_\theta ( {\bm S}_{\theta} G_{P\theta 0}(s) + {\bm L} )^{-1}
			 	{\bm \varPsi}^{-\top}_\theta {\bm \varPsi}^{\top}_\theta \Delta {\bm P}_{Ext} 
 	\vspace{1ex}\\
		\quad  	=  {\bm \varPhi}_\theta  ( {\bm S}_{M\theta} G_{P\theta 0}(s) + {\bm L}_{M\theta})^{-1}
		 		{\bm \varPsi}^{\top}_\theta \Delta {\bm P}_{Ext}
 	% \vspace{1ex}\\
 	% 	\quad 	= \sum\limits_{k=1}^{n} 
 	% 		\dfrac{ {\bm \varphi}_{\theta,k}  {\bm \psi}_{\theta,k}^\top  \Delta {\bm P}_{Ext} } 
	% 		{ {S}_{M\theta,k} G_{P\theta 0}(s) + {L}_{M\theta,k}  }	 		
 	\vspace{1ex}\\
 		\quad 	= \sum\limits_{k=1}^{n} 
 			\dfrac{ \omega_0 {\bm \varphi}_{\theta,k}  {\bm \psi}_{\theta,k}^\top  \Delta {\bm P}_{Ext} } 
			{ {J}_{M\theta,k} s^2 + {D}_{M\theta,k}s  + {K}_{M\theta,k}  } := \sum\limits_{k=1}^{n} \Delta {\bm \theta}_k\,,
	\end{array}
\end{equation}
	where ${J}_{M\theta,k} = {S}_{M\theta,k} J_{P\theta 0}$, ${D}_{M\theta,k} = {S}_{M\theta,k} D_{P\theta 0}$ and ${K}_{M\theta,k} = {S}_{M\theta,k} K_{P\theta 0} + \omega_0 {L}_{M\theta,k}$ are respectively the {\it modal inertia, modal damping and modal spring constant} of the $k$-th frequency mode. Note that modal inertia and modal damping are decided by devices only, while modal spring constant is also related to the network. Here, $\Delta {\bm \theta}_k$ is the $k$-th modal angle component, which can be obtained by the equivalent system shown in Fig.~\ref{Fig_equivalent_system}~(a). Each equivalent system corresponds to a generalized eigenvalue, and the combination of all such systems forms the original system. They are therefore referred to as {\it eigen-subsystems}.

	The modal frequency component $\Delta {\bm \omega}_k$ can be obtained as
\begin{equation} \label{Eq_f_decouple}
	\begin{array}{l}
		\Delta {\bm \omega} =  \dfrac{s}{\omega_0} \Delta {\bm \theta}
	\vspace{1ex}\\
		\;\;\;
		 = \sum\limits_{k=1}^{n} 
			\dfrac{ s {\bm \varphi}_{\theta,k}  {\bm \psi}_{\theta,k}^\top  \Delta {\bm P}_{Ext} } 
				{ {J}_{M\theta,k} s^2 + {D}_{M\theta,k}s  + {K}_{M\theta,k}  }
			:= \sum\limits_{k=1}^{n} \Delta {\bm \omega}_k\,.
	\end{array}
\end{equation}

	Similarly, the voltage response in \eqref{Eq_F_V_response_decoupling} can be decoupled as
\begin{equation} \label{Eq_V_decouple}
	{\bm V}_e^{-1} \Delta {\bm V} 
		= \sum\limits_{k=1}^{n} 
			\dfrac{ {\bm \varphi}_{V,k}  {\bm \psi}_{V,k}^\top  \Delta {\bm  Q}_{Ext} } 
			{ {D}_{MV,k}s + {K}_{MV,k}  }
			:= \sum\limits_{k=1}^{n} {\bm V}_e^{-1} \Delta {\bm V}_k	\,,		
\end{equation}
	where ${D}_{MV,k} = {S}_{MV,k} D_{V  Q 0}$, ${K}_{MV,k} = {S}_{MV,k} K_{V  Q 0} + {L}_{MV,k}$ is the modal damping and spring of the $k$-th voltage mode. Fig.~\ref{Fig_equivalent_system}~(b) shows the corresponding eigen-subsystem. 
	% Note that in this paper, the term ``voltage stiffness'' refers to the ratio of reactive power change to static voltage deviation, which is different from the definition of voltage stiffness in \cite{wang2024voltagestiffness}, where it is defined as the change in bus voltage before and after the device is connected.

\subsection{Global/Local Characteristics of F/V Eigen-subsystems}

\begin{figure}[!t]
    \centering
    \includegraphics[width=0.6\linewidth]{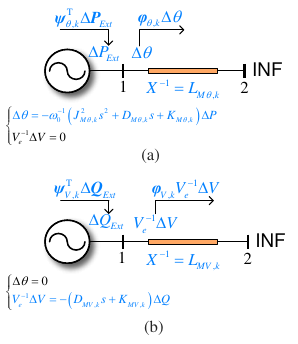}
    \vspace{-3mm}
    \caption{Eigen-subsystems of (a) frequency (angle) and (b) voltage, where $\Delta P_{Ext}$ and $\Delta  Q_{Ext}$ are the equivalent disturbances,  $\Delta\theta$ and $V_e^{-1} \Delta V$ are the responses. Here, $V_{e,1} = V_{e,2} = 1$, and $\theta_{e,12} = 0$.}
    \vspace{-2mm}
    \label{Fig_equivalent_system}
\end{figure}
	
	To further understand the characteristics of the eigen-subsystems, we analyze their global or local properties. This requires examining the properties of the generalized eigenvalues and eigenvectors in \eqref{Eq_generalized_eigenvalue}. They are typically real. Specifically, the term $G_{ij} \sin \theta_{e,ij}$ is significantly smaller than $B_{ij} \cos \theta_{e,ij}$. Neglecting the former renders the $({\bm L}, {\bm S}_{\theta})$ symmetric. Thus, its eigenvalues and eigenvectors are generally real. 
    Then, the generalized eigenvalues of $({\bm L}, {\bm S}_{\theta})$ can be sorted such that $\lambda_{\theta,1} \leq \lambda_{\theta,2} \leq ... \leq \lambda_{\theta,n}$. 
	Moreover, we normalize the eigenvectors by setting their elements with largest module to be 1, as follows, for easier analysis.
\begin{equation} \label{Eq_eigenvector_normalization}
	\max_i |{\varphi}_{\theta,k,i}| = 1, \quad \max_i |{\psi}_{\theta,k,i}| = 1, 
		\quad  \forall k\,.
\end{equation}

	Considering the invariance of power flow \eqref{Eq_invariance}, ${\bm 1}$ is a right eigenvector of $({\bm L}, {\bm S}_{\theta})$, with an eigenvalue of zero. When $S_{\theta,i}, S_{V,i} > 0$ ($\forall i$) and the network is connected, the matrix pencil has only one zero eigenvalue, with all other eigenvalues being positive \cite{dorfler2012kron}. Then, $\lambda_{\theta,1} = \lambda_{V,1} = 0$  and ${\bm \varphi}_{\theta,1} = {\bm \varphi}_{V,1} = {\bm 1}$. From \eqref{Eq_f_decouple} and \eqref{Eq_V_decouple}, the eigenvector ${\bm 1}$ indicates a global component in both F/V response, where all buses experience the same response. These components are termed the CM components, including CM-F and CM-V. It is worth mentioning that $L_{M\theta,1} = L_{MV,1} = 0$, meaning that the line impedance in the CM subsystems in Fig.~\ref{Fig_equivalent_system} is infinite.
	
	The remaining right eigenvectors are approximately weighted orthogonal to ${\bm 1}$, as indicated by the relation ${\bm \varphi}_{\theta,k} \approx {\bm \psi}_{\theta,k}$ and ${\bm \psi}_{\theta,k}^{\top} {\bm S}_{\theta} {\bm 1} = 0$ ($k > 1$), as shown in \eqref{Eq_F_modal_matrix}. These components, which exhibit variations across buses, are referred to as DM components, including DM-Fs and DM-Vs.

	In engineering practice, frequency commonly exhibits global behavior, while voltage is often considered local. This observation seems to contradict the above analysis, where both F/V have their CM components. The reason for this discrepancy is that CM-V is generally negligible.

	Specifically, consider $\Delta {\bm P}_{Ext} = -{\bm P}_0 / s$ and  $\Delta {\bm  Q}_{Ext} = -{\bm  Q}_0 / s$, where ${\bm P}_0$ and ${\bm  Q}_0$ are the disturbance size vectors. Applying the finial value theorem in \eqref{Eq_f_decouple} and \eqref{Eq_V_decouple}, we obtain:
\begin{equation} \label{Eq_f_tinf}
	\left. \Delta {\bm \omega}_k (t) \right|_{t\to \infty} =
		\left\{ \begin{array}{lc}
			- \dfrac{ {\bm \varphi}_{\theta,k}  {\bm \psi}_{\theta,k}^\top  {\bm P}_0 } 
				{ {D}_{M\theta,k} }, &  {K}_{M\theta,k} = 0 \,,
			\vspace{1ex}\\
			0 , & {K}_{M\theta,k} > 0 \,.
	    \end{array} \right.  
\end{equation}
\begin{equation} \label{Eq_V_tinf}
	\left. {\bm V}_e^{-1} \Delta {\bm V}_k (t) \right|_{t\to \infty} =
		- \dfrac{ {\bm \varphi}_{V,k}  {\bm \psi}_{V,k}^\top  {\bm  Q}_0 } 
			{ {K}_{MV,k}  }\,.
\end{equation}

	Eq.~\eqref{Eq_f_tinf} demonstrates that $\Delta {\bm \omega}_k$ approaches zero when ${K}_{M\theta,k} > 0$. It is reasonable as spring acts as an integral control w.r.t. frequency (velocity). In frequency control, devices typically provide negligible or zero spring (SFR), and the network also offers no spring for the CM-F ($\omega_0 {L}_{M\theta,1} =  0$). Consequently, the CM-F spring constant is zero or very small, resulting in a significant CM-F response. In contrast, the network supplies substantial spring for DM-Fs ($\omega_0 {L}_{MV,k} > 0$), which causes DM-Fs to converge rapidly to zero (combined with damping). Therefore, frequency responses are typically dominated by the CM-F, displaying a global characteristic.

	On the other hand, spring is a proportional control w.r.t. voltage (displacement), which inherently results in steady-state deviations, as shown in \eqref{Eq_V_tinf}. For the CM-V, the modal spring is determined by the sum of all devices' contributions (it will further be explained in Section IV-D).
	% (${K}_{MV,1} = {S}_{MV,1} K_{V  Q 0}$, where ${S}_{MV,1} = {\bm \psi}_{V,1}^{\top} {\bm S}_{V} {\bm \varphi}_{V,1}$ and ${\bm \psi}_{V,1} \approx {\bm \varphi}_{V,1} = {\bm 1}$). 
	Many devices, such as SGs, provide large voltage spring constant. Thus, this cumulative spring is typically very large, leading to a negligible CM-V response. Conversely, for the DM-V, while the network provides ${L}_{MV,k} > 0$, the contributions from devices are no longer cumulative and significantly reduced. Thus, DM-V responses tend to be larger than CM-V, and the voltage exhibits a local characteristic.

\begin{remark}
	The above analysis shows that frequency typically exhibits global behavior, while voltage tends to be local. The reason behind is that {\it frequency corresponds to a velocity term, while voltage corresponds to a displacement term}. 
	However, their spatial distribution is not always as such. For instance, in a single-device infinite-bus system, the CM-F is zero, and the frequency response is entirely the DM-F, as demonstrated in Case 1-c in Section V-B. On the other hand, in a PE-denominated power system, if the IBRs in grid-forming mode reach their capacity limits and fail to form the voltage, the CM-V spring constant will significantly decrease, which may lead to CM-V instability, as illustrated in Section VI-C.
\end{remark}

\subsection{Corresponding Modal Power Components}
	
	The active power response corresponding to the each frequency component can be expressed as 
\begin{equation} \label{Eq_rho_decouple}
	\begin{array}{l}
		\Delta {\bm P} = - {\bm S}_{\theta} G_{P\theta}(s) \Delta {\bm \theta}
 	\vspace{1ex}\\
 		\;\; 	= -\sum\limits_{k=1}^{n} 
 			\dfrac{ {\bm S}_{\theta} {\bm \varphi}_{\theta,k}  {\bm \psi}_{\theta,k}^\top  G_{P\theta 0}(s) \Delta {\bm P}_{Ext} } 
				{ {S}_{M\theta,k} G_{P\theta 0}(s) + {L}_{M\theta,k}  } := \sum\limits_{k=1}^{n} \Delta {\bm P}_k		\,,
	\end{array}
\end{equation}
% \begin{equation} \label{Eq_sigma_decouple}
% 	\begin{array}{l}
% 		\Delta {\bm  Q} = - {\bm S}_{V} G_{V  Q}(s)  {\bm V}_e^{-1} \Delta {\bm V}
%  	\vspace{1ex}\\
%  		\;\;	= -\sum\limits_{k=1}^{n} 
%  			\dfrac{ {\bm S}_{V} {\bm \varphi}_{V,k}  {\bm \psi}_{V,k}^\top  G_{V  Q 0}(s) \Delta {\bm  Q}_{Ext} }
% 				{ {S}_{MV,k} G_{V  Q 0}(s) + {L}_{MV,k}  } := \sum\limits_{k=1}^{n} \Delta {\bm  Q}_k		\,.
% 	\end{array}
% \end{equation}	 
	Left-multiplying \eqref{Eq_rho_decouple} with ${\bm \psi}_{\theta,1}^\top$, we obtain
\begin{equation} \label{Eq_rho_sum}
{\bm \psi}_{\theta,1}^\top \Delta {\bm P}_k = \left\{
	\begin{array}{lc}
		{\bm \psi}_{\theta,1}^\top \Delta {\bm P}_{Ext} , &  k = 1 \,,
		\vspace{1ex}\\
		0 , & k > 1 \,.
    \end{array}
    \right. 
\end{equation}
% \begin{equation} \label{Eq_sigma_sum}
% {\bm \psi}_{V,1}^\top \Delta {\bm  Q}_k = \left\{
% 	\begin{array}{lc}
% 		{\bm \psi}_{V,1}^\top \Delta {\bm  Q}_{Ext} , &  k = 1 \,,
% 		\vspace{1ex} \\
% 		0 , & k > 1\,.
%     \end{array}
%     \right. 
% \end{equation}
	
	Combined with \({L}_{M\theta,1} = 0\), \eqref{Eq_rho_decouple}-\eqref{Eq_rho_sum} indicate that the CM power of each device is a share of the disturbance (\({\bm \psi}_{\theta,1}^\top \Delta {\bm P}_{Ext}\)), weighted by their respective capacities (\({\psi}_{\theta,1,i}{S}_{\theta,i} / {S}_{M\theta,1}\)). The sum of these contributions equals the total disturbance. This implies that the CM-F, along with their corresponding power responses, reflect the collective ability of all devices to absorb external disturbances. 

	In contrast, the DM power response is influenced by not only device parameters but also the disturbance location and the network topology. The DM response fills the gap between the actual power output of devices and the CM component. Importantly, the sum of all devices' DM power is zero, which indicates that they represent a redistribution of the disturbance power within the system. 

	The above analysis uses frequency as an example, but it also applies to voltage. A minor distinction lies in that, the reactive power dynamics of devices ($G'_{VQ\,0}(s)$) includes the term $2Q_e$, which does not inherently belong to the device. Therefore, the reactive power response derived similarly to \eqref{Eq_rho_decouple} is ``virtual''. It is worth noting that this virtual power closely approximates the actual power when the device's spring constant $K_{QV}$ is significantly larger than $2Q_e$.

	% A distinction between the DM-F and DM-V is that the former decays over time, while the latter does not. Consequently, the DM active power response converges to zero over time (in the absence of device stiffness), and the active power response of each device eventually converges to the CM component. In contrast, the reactive power responses of devices continue to include both the CM and DM components.

	% In frequency analysis, a traditional view is that (active) power disturbance is first distributed to devices based on their electrical distance from the disturbance point, followed by distribution according to the device inertia, and finally by damping or PFR. However, our analysis reveals that this distribution can be more accurately represented by two concurrent processes: one based on inertia and PFR ($S_{\theta,i}$), corresponding to CM power; and the other based on electrical distance, corresponding to DM power.

\section{F/V Strength Quantification based on \\ Eigen- subsystems}

	Based on the eigen-subsystems, this section proposes F/V strength metrics, and compares them with traditional metrics.

\subsection{Strength Quantification Approach}

	Literally, frequency and voltage strength refer to the robustness of a power system's F/V responses—that is, the system’s ability to withstand disturbances and maintain F/V stability. Accordingly, the magnitude of the F/V response under a given disturbance can be used to quantify strength, as expressed in \eqref{Eq_strength_definition}. The smaller the response, the stronger the system.
\begin{equation} \label{Eq_strength_definition}
	\frac{\| \Delta {\bm \omega} \|} {\| \Delta {\bm P_{Ext}}\|} 
	\Rightarrow \frac{\| \Delta {\bm \omega}_k \|} {\| \Delta {\bm P_{Ext}}\|},  
	\frac{\| {\bm V}_e^{-1} \Delta {\bm V} \|} {\| \Delta {\bm Q_{Ext}}\|}
	\Rightarrow \frac{\| {\bm V}_e^{-1} \Delta {\bm V}_k \|} {\| \Delta {\bm Q_{Ext}}\|}.
\end{equation}
The ratio between the response and disturbance vectors is directly related to singular value or $\mathcal{H}_\infty$-norm. However, such metrics are typically obtained through numerical algorithms and are difficult to interpret in terms of the physical parameters of the original system. Furthermore, the overall system response comprises various modal components, each exhibiting distinct response characteristics. To quantify strength with clearer physical meaning, we instead utilize the parameters of each eigen-subsystem as metrics, as detailed below.

\subsection{Strength Quantification of CM-F}

	Given the eigenvector normalization method \eqref{Eq_eigenvector_normalization}, the modal parameters ${J}_{M\theta,k}$, ${D}_{M\theta,k}$, and ${K}_{M\theta,k}$ quantify the ability of each eigen-subsystem to withstand a unit disturbance applied at the most sensitive bus, which induces the largest response. These parameters can therefore serve as modal frequency strength metrics. Physically, modal inertia relates to the initial RoCoF; modal damping governs the steady-state deviation (in the absence of modal spring constant); and modal spring constant determines the speed of frequency recovery.

	Due to the approximate symmetry of $({\bm L}, {\bm S}_{\theta})$, it holds that ${\bm \psi}_{\theta,1} \approx {\bm \varphi}_{\theta,1} = {\bm 1}$. This implies that 1) the three CM parameters are obtained by summing the parameters of all devices  \eqref{Eq_F_modal_matrix}, and 2) they characterize the CM-F response of all buses under any single-point disturbance. However, when the network losses are non-negligible and the symmetry is broken, ${\bm \psi}_{\theta,1} \neq {\bm 1}$, e.g, in systems with substantial constant impedance loads \cite{gao2022common}. In such cases, the CM-F response depends on disturbance location, and bus-specific strength metrics are required. This will be discussed in the following.

\subsection{Strength Quantification of DM-F}

	For DM-F, the responses at different buses vary significantly, and the impact of disturbances applied at different locations also differs. Therefore, bus-specific strength metrics are required. When a disturbance is applied at bus $j$ and  the response is observed at bus $i$,  the corresponding bus-specific modal metrics can be defined as:
\begin{equation} \label{Eq_bus_modal_parameters}
\left\{
	\begin{array}{ll}
		{J}_{M\theta,k}^{(i,j)} = \dfrac{ {J}_{M\theta,k} } { {\varphi}_{\theta,k,i}  {\psi}_{\theta,k,j}} ,
		& {D}_{M\theta,k}^{(i,j)} = \dfrac{ {D}_{M\theta,k} } { {\varphi}_{\theta,k,i}  {\psi}_{\theta,k,j}} ,
		\vspace{1ex} \\
		{K}_{M\theta,k}^{(i,j)} = \dfrac{ {K}_{M\theta,k} } { {\varphi}_{\theta,k,i}  {\psi}_{\theta,k,j}} \,. &
    \end{array}
\right. 
\end{equation}
% where superscript $(i,j)$ indicates the bus at which the response is observed and the bus where the disturbance is applied. 
Note that these bus-specific metrics can also apply to CM-F when disturbance position matters.

\subsection{Strength Quantification of CM-V}
	
	Similar to frequency strength, voltage strength can be quantified using modal parameters, i.e., ${D}_{MV,k}$ and ${K}_{MV,k}$. Among these, modal spring constant $K_{MV,k}$ is a critical metric, which is directly related to the static voltage stability. Smaller modal spring indicates greater voltage deviations under disturbances, as shown in \eqref{Eq_V_tinf}. Moreover, $K_{MV,k}$ approaching zero indicates the risk of static voltage collapse. 

	Note that, unlike frequency-related parameters (e.g., $J_{P\theta}$) which are typically positive, voltage-related parameters may be negative. For example, the $G_{QV}(s)$ of constant reactive power loads (CRPLs) is $G_{QV}(s) =  -2Q_e < 0$. In systems that includes both CRPLs and generation devices, their voltage dynamics are heterogeneous, making direct decoupling with method in \eqref{Eq_theta_decouple} infeasible. Decoupling such heterogeneous systems will be a direction for future work. In this paper, we adopt a simplified static analysis (namely, set $s = 0$ and focus on the spring term of devices) to bypass this problem.

	Let the spring constants of generation devices and CRPLs be denoted by ${\bm K}_{QV} = \mathrm{diag}\{ K_{QV,i} \}$ and $-2{\bm Q}_e = -\mathrm{diag}\{ 2Q_{e,i} \}$, respectively, and $G_{QV 0} = 1$. Then, ${\bm S}_{V} = \text{diag} \{ {\bm K}_{QV}, - 2{\bm Q}_e\}$. Given that ${\bm \psi}_{1} \approx {\bm \varphi}_{1} = {\bm 1}$,  the modal spring constant of CM-V is 
\begin{equation} \label{Eq_CM_V_spring}
	K_{MV,1} \approx \sum\nolimits_i  {K}_{VQ,i} -  \sum\nolimits_i 2Q_{e,i} \, ,
\end{equation}
	which means, the CM-V is stable when the total positive spring constant of generation devices is larger than the total negetive spring constant of loads.

	% Considering $G_{QV 0} = 1$, then $S_{V,i}$ represent the stiffness of the devices or CRPLs, which is positive for generation devices and negative for CRPLs.

\subsection{Strength Quantification of DM-V}

	The DM-V strength can also be quantified by modal parameters, especially $K_{MV,k}$. Meanwhile, bus-specific metrics can be defined like \eqref{Eq_bus_modal_parameters}. 

	It is worth mentioning that the DM-V spring constant is closely related to the gSCR metric. To illustration this connection, consider the characteristic equation of the system
\begin{equation} \label{Eq_det_V}
		\lambda_{V,k} 
		\begin{bmatrix} {\bm K}_{QV} & \\ & - 2{\bm Q}_e \end{bmatrix} 
		\begin{bmatrix} {\bm \varphi}_{k}^{\{ 1\}} \\ {\bm \varphi}_{k}^{\{ 2\}} \end{bmatrix} 
	=
		\begin{bmatrix} {\bm L}_{11} & {\bm L}_{12} \\ {\bm L}_{21} & {\bm L}_{22} \end{bmatrix}
		\begin{bmatrix} {\bm \varphi}_{k}^{\{ 1\}} \\ {\bm \varphi}_{k}^{\{ 2\}} \end{bmatrix}
\end{equation}
	where the superscript indicates generators (${\{ 1\}}$) or loads (${\{ 2\}}$).

	Analyses based on the gSCR or other SCR-related metrics typically assume the existence of infinite buses, i.e., ${\bm K}_{QV} \to \infty$. Under this assumption, the eigenvectors satisfy ${\bm \varphi}_{k}^{\{1\}} \to {\bm 0}$ for DM-V ($\lambda_{V,k} \neq 0$). Consequently, the system reduces to $({\bm L}_{22}, -2{\bm Q}_e)$, i.e.,
\begin{equation} \label{Eq_det_V_red}
		\lambda_{V,k} (- 2{\bm Q}_e) {\bm \varphi}_{k}^{\{ 2\}}
	=
		{\bm L}_{22} {\bm \varphi}_{k}^{\{ 2\}} \, .
\end{equation}

	For system $({\bm L}_{22}, -2{\bm Q}_e)$, gSCR is defined as the smallest eigenvalue of ${\bm Q}_e^{-1}{\bm L}_{22}$ \cite{zhou2023hetero}, which equals to $-2\lambda_{V,k}$ ($\lambda_{V,k}$ is the smallest generalized eigenvalue in \eqref{Eq_det_V_red}). Substituting this relationship into the modal spring constant calculation yields:
\begin{equation} \label{Eq_K_gSCR}
	{K}_{MV,k} = 
		(gSCR -2) {\bm \psi}^{\{ 2\}}_{k} {\bm L}_{22} {\bm \varphi}_{k}^{\{ 2\}}\,.
\end{equation}
	It indicates that gSCR below 2 is equivalent to ${K}_{MV,k} < 0$, both of which indicate static voltage collapse. Note that this threshold 2 comes from the load ($- 2{\bm Q}_e$).
	 %Here, detailed derivations are omitted due to page limitation.

\begin{remark}
	In static voltage stability analysis, the gSCR (as well as other SCR-based metrics) focuses on the DM-V associated with the smallest nonzero generalized eigenvalue $\lambda_{V,k}$, which can be referred to as the first DM-V. These analyses typically assume that the loads (or HVDC) exhibit similar dynamics. Under such condition, the first DM-V is generally the most vulnerable to instability \cite{xin2016gSCR_PED}. However, it is worth noting that in systems with more complex and heterogeneous device dynamics, other DM-V modes may also become critical, which deserves further investigation. 
\end{remark}

\subsection{Comparisons of F/V Strength Metrics}

	For a comprehensive understanding, we compare the proposed F/V strength metrics with existing ones. Furthermore, the CM and DM metrics are compared to demonstrates their differences.

\subsubsection{Frequency Strength} 

	Traditionally, the total inertia is taken as a frequency strength metric. Besides, recent studies introduce nodal inertia ($J^{(i,j)}$), defined as the disturbance size at bus $j$ divided by the initial RoCoF at bus $i$. These metrics, however, might have certain limitations. 
	Specifically, a) the total inertia is essentially the same as the CM inertia ${J}_{M\theta,1}$, and does not account for DM-F;
	b) nodal inertia aggregates the contribution of all modes in one bus, masking individual modal contributions and prevents detailed analysis. Note that the initial RoCoF of a bus is the sum of all modal RoCoFs:
\begin{equation} \label{Eq_bus_inertia_modal_inertia}
	\left( J^{(i,j)} \right)^{-1} = \sum\nolimits_k \left( {J}^{(i,j)}_{M\theta,k} \right)^{-1}\,.
\end{equation}
	
	In contrast, the proposed metrics, which capture mode-specific characteristics, address these limitations, which will be further demonstrated in the next section.

\subsubsection{Voltage Strength} 

	Conventional voltage strength metrics are based on SCR, which typically assume the presence of infinite buses. This assumption is reasonable in traditional power systems, where SGs possess strong overload capabilities and can effectively maintain their terminal voltages. In such cases, the evaluation of DM-V static stability based on modal spring constant is equivalent to that based on the gSCR, as illustrated in \eqref{Eq_K_gSCR}.

	However, in PE-dominated power systems, this assumption may no longer hold. The IBRs, which operate as voltage sources (e.g., in grid-forming mode), are likely to reach their capacity limits under disturbances. As a result, their ${K}_{QV}$ significantly decreases, and treating them as infinite buses can lead to overly optimistic evaluations for DM-V. Moreover, in such cases, the CM-V response may no longer be negligible, and its strength need be assessed. In this context, the proposed voltage metric of modal spring constant, which enables the evaluation of both CM-V and DM-V, is more comprehensive than the SCR-based metrics.

	A limitation of the proposed metrics is that, they focus on low-frequency responses, given the use of a static network model. In contrast, metrics such as gSCR can also capture small-signal stability characteristics in higher frequency ranges. Extending the proposed method to account for higher frequency dynamics will be our future research direction.

\subsubsection{CM and DM}

	Although similar metrics are used to assess CM and DM components, they reflect fundamentally different aspects of the capability of power systems. Specifically, CM metrics are determined by the system’s total power support capacity, whereas DM metrics are more closely linked to the power exchange capability between different areas, as indicated in \eqref{Eq_rho_sum}. As a result, CM metrics depend primarily on the characteristics of the devices, while DM metrics are influenced by not only the devices but also the network.

\section{Illustrative Examples}

\begin{figure}[!t]
    \centering
    \includegraphics[width=0.7\linewidth]{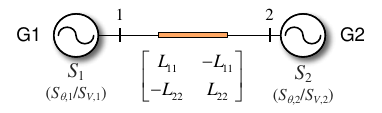}
    \vspace{-3mm}
    \caption{Demonstration of the two-device test system.}
    \vspace{-2mm}
    \label{Fig_2G}
\end{figure}

	This section demonstrates the modal decoupling through a two-device system, depicted in Fig.~\ref{Fig_2G}, for which we can derive closed-form solutions to cross-verify the numerical results.

\subsection{Closed-form Solutions}

	For simplicity, we represent the system using the following notations: ${\bm x}$ denotes the angle or voltage responses, ${\bm y}_{Ext}$ denotes the external power disturbance, ${\bm S} = \text{diag}\{ S_1, S_2 \}$ denotes the device ratios, $G(s)$ denotes the nominal dynamics of devices, and ${\bm L} = [L_{11}, -L_{11}; -L_{22}, L_{22}]$ is the network matrix. Besides, it is assumed that $V_{e,1} = V_{e,2} = 1$.

	Using the method in Section III, ${\bm x}$ can be decoupled as:
\begin{equation} \label{Eq_decouple_2G_symbolic}
\begin{array}{l}
	{\bm x} 
	= 
	% \vspace{1ex} \\
	\left(
	\dfrac{ 	\begin{bmatrix} 1 \\ 1 \end{bmatrix} 
				\begin{bmatrix} \dfrac{L_{22}}{L_{11}} & 1 \end{bmatrix} }
		{ L_{11}^{-1} S_{L}  G (s) } 
	% {\bm x}
	% \quad + 
	+
	\dfrac{ 	\begin{bmatrix} -\dfrac{L_{11}S_2}{L_{22}S_1} \\ 1 \end{bmatrix} 
				\begin{bmatrix} -\dfrac{S_2}{S_1} & 1 \end{bmatrix} }
		{ \dfrac{S_{L} S_2} {L_{22} S_1} G (s)  +  \dfrac{S_{L}^2} {L_{22} S_1^2} } 
	\right) 
	{\bm y}_{Ext} \,,
\end{array}
\end{equation}
where $S_{L} = L_{22}S_1 + L_{11}S_2$. 

	According to \eqref{Eq_decouple_2G_symbolic}, we conclude that 1) for a symmetric network ($L_{11} = L_{22}$), the CM response depends solely on the aggregate devices ($L_{11}^{-1}S_L = S_1 + S_2$) and disturbances in the system, not their locations or the network structure. In contrast, asymmetric networks introduce different weightings for disturbances and devices at different buses, 
	% For example, if $L_{22}/L_{11} = 1.1$, disturbance at Bus 1 result in a response 1.1 times stronger than that at Bus 2, 
	as also discussed in \cite{gao2022common}.
	2) DM responses are heavily influenced by the device and disturbance locations. For $S_1, S_2 > 0$, elements in DM eigenvector have opposite signs, indicating opposite response of two buses, such as the relative oscillation of DM-F. For $S_1$ and $S_2$ with opposite signs, the responses of two buses share the same sign, e.g., in the collapse of DM-V shown in Fig.~\ref{Fig_simu_voltage_collapse}. 

	Next, we analyze the proposed strength metrics. For simplicity, a symmetric network $L_{11} = L_{22} = L$ is considered.
	For frequency strength, let $J_1$, $J_2$ denote the inertia of G1 and G2, respective. Then, the two modal inertia of bus 1 are 
\begin{equation}
	{J}^{(1,1)}_{M\theta,1} = J_1 + J_2, \quad
		{J}^{(1,1)}_{M\theta,2} = \dfrac{(S_1 + S_2) J_1} {S_2}\,.
\end{equation}
Obviously, the CM inertia is the total inertia. Besides, we have $({J}^{(1,1)}_{M\theta,1} )^{-1} + ({J}^{(1,1)}_{M\theta,2} )^{-1} = (J_1)^{-1}$, which means that the nodal inertia $J_1$ is decided by two modal inertia, as in \eqref{Eq_bus_inertia_modal_inertia}.

	For voltage strength, consider G2 as a CRPL and let $G_{QV0}(s) = 1$, then $S_1 > 0 > S_2$. The two modal spring constants are:
\begin{equation} \label{Eq_Voltage_Strength_Example}
\left\{
	\begin{array}{l}
		{K}_{MV,1} = S_{sum}\,,
	\vspace{1ex}\\
		{K}_{MV,2} =  S_1^{-2} S_{sum} (S_1 S_2 + S_{sum} L)\,,
    \end{array}
\right. 
\end{equation}
	where $S_{sum} = S_1 + S_2$. From \eqref{Eq_Voltage_Strength_Example}, the CM-V collapse occurs when $S_{sum} \leq 0$, indicating total spring of devices is non-positive. Note that it is independent of the network. On the other hand, the condition of DM-V collapse is equivalent to 
\begin{equation} \label{Eq_L}
	S_2 \leq \dfrac{-L} { S_1^{-1} L + 1 } := S_{2,th} \,,
\end{equation}
	and $S_{sum} > 0$.
	It indicates that the stability threshold ($S_{2,th}$) depends on both device $S_1$ and network $L$. As $S_1$ increases (the support of G1 increases), the threshold for $S_2$ becomes less stringent. When $S_1 \to \infty$, we have $S_{2,th} = -L$.

\subsection{Numerical Results}

	In the following, we present numerical examples to illustrate the F/V modal components in a two-device test system.

\subsubsection{Case 1, Modal Frequency Analysis}

	The system is analyzed with $G_{P\theta 0}(s) = (Js^2 + Ds)/\omega_0$ and ${\bm L} = [3, -3; -3, 3]$. Table~\ref{Tab_para_2G_f} summarizes the parameters used for Case 1. 

\begin{figure}[!t]
    \centering
    \includegraphics[width=0.9\linewidth]{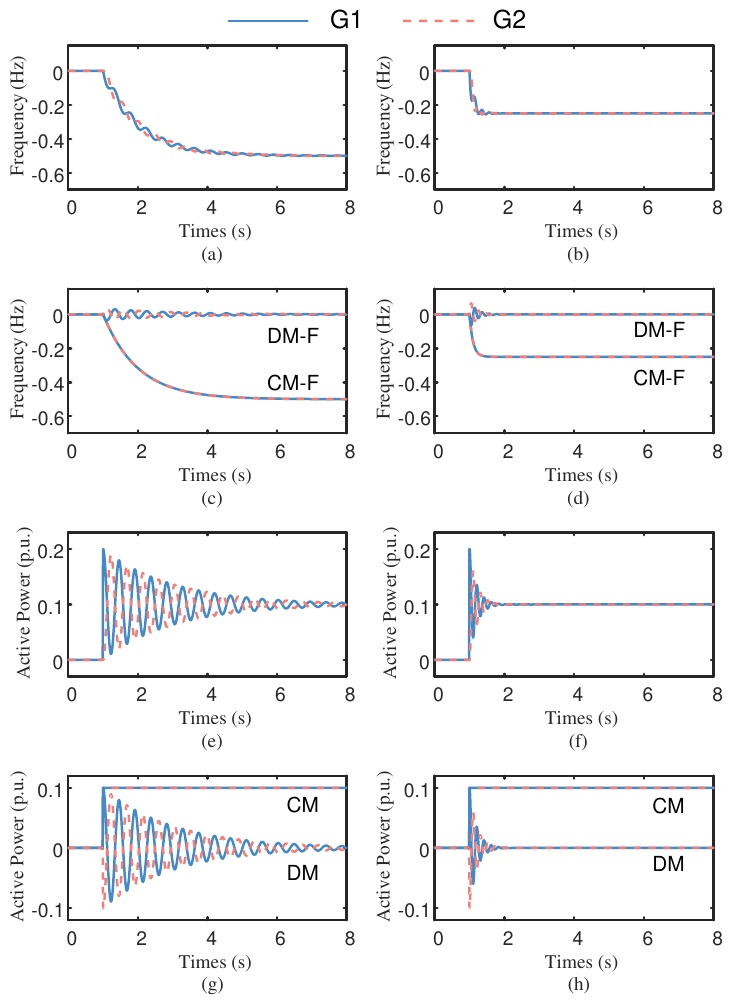}
    \vspace{-3mm}
	\caption{Frequency responses $\Delta\omega_i$ (a, b) and active power responses $\Delta P_i$ (e, f) along with their corresponding modal components (c, d and g, h) in Cases~1-a (left side) and 1-b (right side).}
    \vspace{-2mm}
    \label{Fig_2G_F}
\end{figure}

	An active power disturbance $\Delta P_{Ext} = -0.2/s$ is applied to G1 at $t = 1$~s. The frequency and active power responses for Cases 1-a and 1-b are shown in Fig.~\ref{Fig_2G_F}. These results demonstrate that increasing damping significantly reduces the CM-F deviation (Fig.~\ref{Fig_2G_F} (c) vs. (d)), while the CM power responses remain identical (Fig.~\ref{Fig_2G_F} (g) vs. (h)). This consistency in CM power responses arises from the equal device inertia and damping in both cases, ensuring an even distribution of the disturbance power between the devices.    
	On the other hand, higher damping and lower inertia also lead to quicker suppression of DM-Fs and the corresponding power oscillations \cite{gao2025inverse}. 

\begin{table}[!t]
		\centering
		\caption{Parameters of Devices and Modal Inertia of Bus 1 in Case 1} 
		\vspace{-2mm}
		\setlength{\tabcolsep}{2.8mm}
		% \resizebox{\columnwidth}{!}{
		\begin{tabular}{ *{6}{c} } 
		\toprule[0.5mm]
		\midrule
				Cases & $[J_{0}, D_{0}]$ & $[S_{1}, S_{2}]$ 
				& $J_1$ & ${J}^{(1,1)}_{M\theta,1}$ & ${J}^{(1,1)}_{M\theta,2}$ \\ 
		\midrule
				1-a & [10, 10] & [1, 1] & 10 & 20 & 20 \\
				1-b & [2, 20] & [1, 1] & 2 & 4 & 4 \\
				1-c & [10, 10] & $[1, \infty]$ & 10 & $\infty$ & 10 \\
				1-d & [10, 10] & [1, 0] & 10 & 10 & \textbackslash  \\
		\bottomrule[0.5mm]
		\end{tabular}
		% }
		\label{Tab_para_2G_f}
\end{table}

	This suggests that 1) with adequate damping, high inertia may not be essential, as it may have little impact on improving the CM-F nadir, reduce the damping ratio of DM-F oscillation, and delay the convergence of device active power outputs to their CM components; 2) in PE-dominated systems, frequency stability (regarding the global frequency) can be greatly enhanced by appropriately increasing the frequency support parameters of all IBRs together. %Consequently, the critical issue in such systems may not be frequency stability but ensuring sufficient reserves of power and energy.  
    Note that, in systems with a lot of SGs, increasing the inertia or damping of IBRs will transfer more disturbance power to them. This could lead to underutilization of the available reserves in SGs.  

\begin{figure}[!t]
    \centering
    \includegraphics[width=0.6\linewidth]{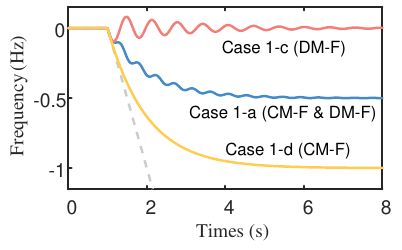}
    \vspace{-3mm}
	\caption{Frequency responses of G1 in Case 1-a, 1-c and 1-d.}
    \vspace{-2mm}
    \label{Fig_modal_inertia}
\end{figure}

	To further emphasize the importance of distinguishing frequency modal components, we compare Cases 1-a, 1-c, and 1-d. Case 1-c represents a single-device infinite bus system, Case 1-d can be viewed as a single-device load system. The frequency responses of G1 under same disturbances in these cases are shown in Fig.~\ref{Fig_modal_inertia}.
	Obviously, Case 1-a includes both CM-F and DM-F, while Case 1-c and 1-d only contains DM-F or CM-F, respectively. Their frequency responses differ significantly, however, the nodal inertia of G1 is identical in the three scenarios. Moreover, the total inertia cannot describe the DM-F in Case 1-a and 1-c. 
	These results underscore the limitations of total inertia and nodal inertia. Instead, modal inertia, as provided in Table~\ref{Tab_para_2G_f}, offers a more precise and insightful measure of frequency strength.

\begin{figure}[!t]
    \centering
    \includegraphics[width=0.9\linewidth]{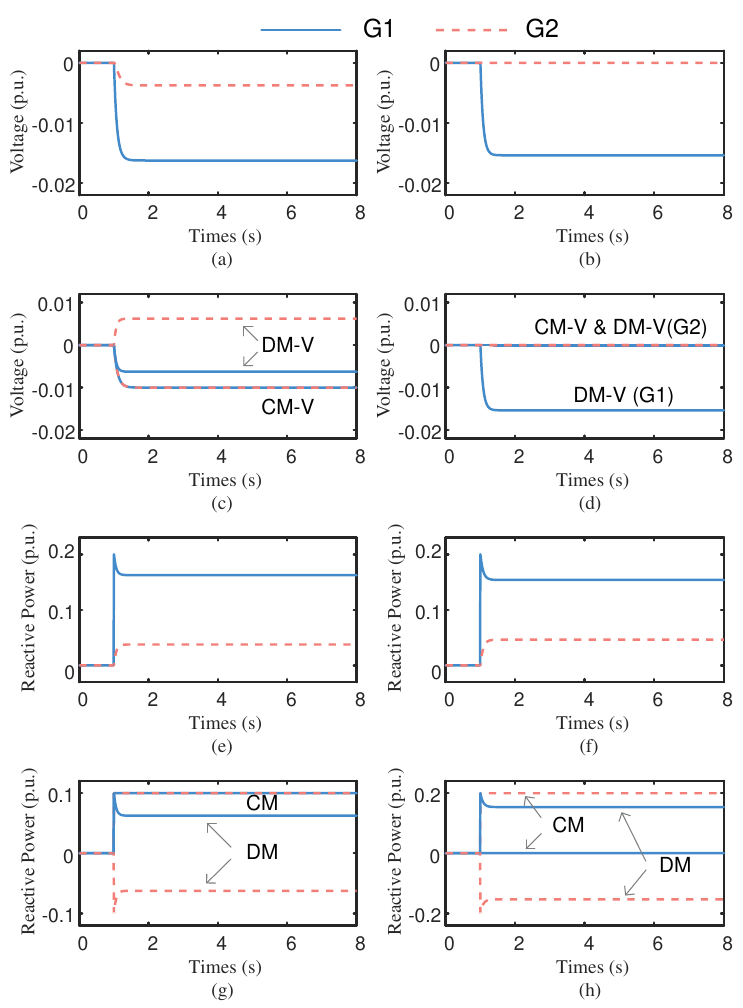}
    \vspace{-3mm}
	\caption{Voltage responses $\Delta V_i / V_{e,i}$ (a, b) and  reactive power responses $\Delta Q_i$ (e, f) along with their corresponding modal components (c, d and g, h) in Cases~2-a (left side) and 2-b (right side).}
    \vspace{-2mm}
    \label{Fig_2G_V}
\end{figure}
	
\subsubsection{Case 2, Modal Voltage Analysis}

	We now analyze the voltage response using the same system as before. The nominal voltage dynamics is $G_{QV 0} = s + 10$. Two scenarios are considered, Case 2-a: $S_1 = S_2 = 1$ and Case 2-b: $S_1 = 1, S_2 = \infty$. In both cases, a reactive power disturbance of 0.2 p.u. is applied to G1. The results are shown in Fig.~\ref{Fig_2G_V}. 
	Fig.~\ref{Fig_2G_V} (c) and (d) reveal that CM-V is generally small, especially when one device provides strong voltage support ($S_2 = \infty$ in Case 2-b). Unlike DM-Fs, DM-Vs and the corresponding power response (Fig.~\ref{Fig_2G_V} (g) and (h))  do not decay over time. They result in voltage responses exhibiting persistent spatial deviations.
	The simplified model does not adequately capture voltage collapse dynamics. It will be illustrated in the next section.

% \section{Modal Components in Complicated Situations}

\begin{figure}[!t]
    \centering
    \includegraphics[width=0.95\linewidth]{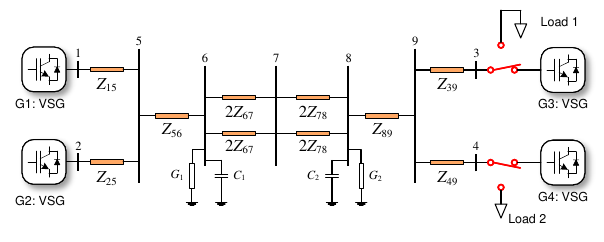}
    \vspace{-3mm}
	\caption{Demonstration of the four-device test system. In Case 3, the switches connect to the VSG, while in Case 4, the switches connect to CRPL.}
    \vspace{-2mm}
    \label{Fig_test_system}
\end{figure}

\begin{figure}[!t]
    \centering
    \includegraphics[width=0.85\linewidth]{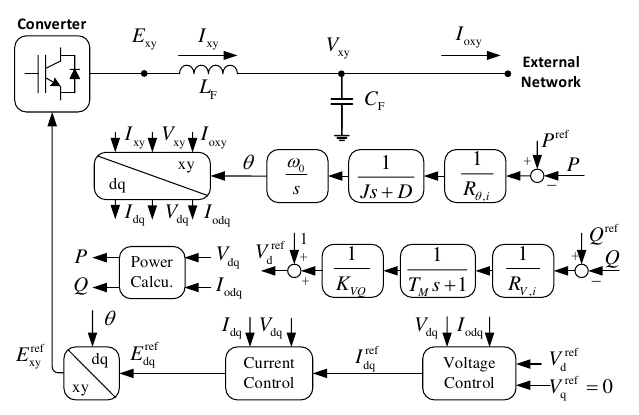}
    \vspace{-3mm}
	\caption{Model of VSGs in simulations.}
    \vspace{-2mm}
    \label{Fig_gen_model}
\end{figure}
	
\section{Time-domain Simulations}

	This section presents time-domain simulations using high-fidelity models to validate the effectiveness of the proposed modal decoupling method and  metrics.

\subsection{Description of the Test System}

	The test system consists of four devices as depicted in Fig.~\ref{Fig_test_system}, with two scenarios. In Case 3, all devices are VSGs, as shown in Fig.~\ref{Fig_gen_model}. This scenario is used to examine the effectiveness of the modal decoupling method. In Case 4, two VSGs are replaced by CRPLs ($Q_{e,3}=Q_{e,4}=0.4$ p.u.), to verify the proposed voltage modal spring metric. The main parameters of four VSGs are: $J = 10$ p.u., $D = 10$ p.u., ${\bm S}_{\theta} = \text{diag}\{ 1,1,2,1 \}$; $K_{QV} = 10$ p.u., $T_{M} = 0.5$~s, ${\bm S}_{V} = \text{diag}\{ 1,1,1,10\}$. Additional parameters and details of the voltage/current controllers for the VSGs can be found in \cite{gao2025inverse}. Line parameters are: $X_{15} = X_{25} = X_{39} = X_{49} = 0.1$ p.u., $X_{56} + X_{67} + X_{78} + X_{89} = 0.15$~p.u. All lines share the same impedance ratio $R/X = 0.1$. $G_1 = G_2 = 0.5$~p.u., $C_1 = C_2 = 0.05$ p.u.

\subsection{Case 3: F/V Modal Decoupling}
	
\begin{figure}[!t]
    \centering
    \includegraphics[width=0.9\linewidth]{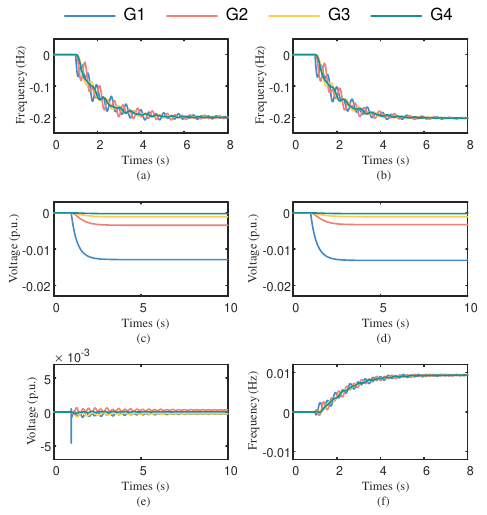}
    \vspace{-3mm}
	\caption{Comparison of simulated and theoretical F/V responses in Case 3. (a) Simulated frequency responses $\Delta\omega_i$, (b) their theoretical results and (e) simulated voltage responses under active power disturbances. (c) Simulated voltage responses $\Delta V_i/V_{e,i}$, (d) their theoretical results and (f) simulated frequency responses under reactive power disturbances.}
    \vspace{-2mm}
    \label{Fig_simu_PQExt}
\end{figure}

	An active power disturbance of 0.2 p.u. is applied at G1, the frequency trajectories of four VSGs are depicted in Fig.~\ref{Fig_simu_PQExt}~(a). These trajectories closely match the theoretical results in Fig.~\ref{Fig_simu_PQExt} (b), which were derived by superimposing four modal components. Meanwhile, voltage responses in this scenario are negligible (Fig.~\ref{Fig_simu_PQExt} (e)), indicating the approximate decoupling of frequency and voltage.

	Similarly, applying a reactive power disturbance at G1 yields voltage responses that align well with theoretical predictions, as shown in Figs.~\ref{Fig_simu_PQExt} (c) and (d). And the frequency response in this case is negligible, shown in Fig.~\ref{Fig_simu_PQExt} (f).  
	These results verify the validity of the proposed modal decomposition approach for both frequency and voltage.

\subsection{Case 4: Static Voltage Collapse}
	
	In this case, two VSGs are replaced with CRPLs to validate the proposed voltage metric. Here, $C_2$ in bus 8 is increased to $0.9$ p.u., to support the bus voltage near the load.

\begin{figure}[!t]
    \centering
    \includegraphics[width=0.9\linewidth]{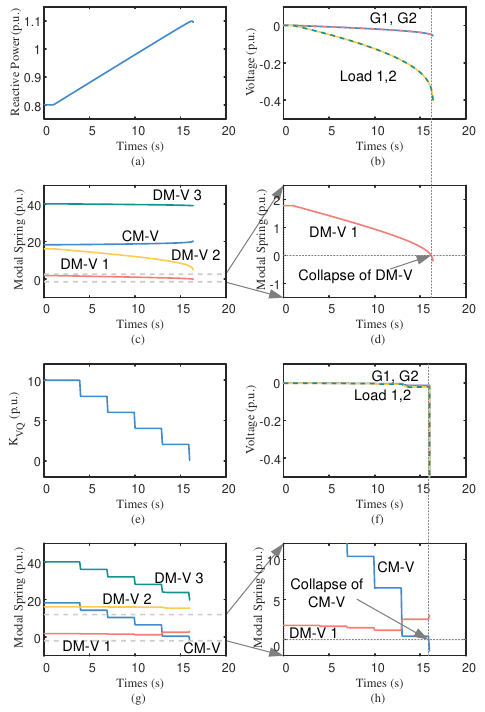}
    \vspace{-3mm}
	\caption{Simulation results in Case 4-a (a)-(d) and Case 4-b (c)-(d). (a): Total reactive power consumed by Load1 and Load2. (e): Q-V droop coefficient $K_{QV}$.  (b) and (f): Voltage trajectories $\Delta V/V_{e}$ of G1, G2, Load1 and Load2. (c) and (g): Modal spring of four voltage mode. (d) and (h): partial enlarged drawing of (c) and (g). 
	}
    \vspace{-2mm}
    \label{Fig_simu_voltage_collapse}
\end{figure}

	First, we gradually increase the reactive power of load (Case 4-a), which leads the spring constant of DM1 to decrease. When it reaches zero, the DM-V voltage collapse occurs, where the load voltage drops much more than the VSG voltage. This process is illustrated in Fig.~\ref{Fig_simu_voltage_collapse} (a)–(d). 

	On the other hand, reducing $K_{QV}$ of the remaining two VSGs (Case 4-b) decreases the spring constant of both DM1 and CM. When $K_{QV}$ exceeds a critical threshold, CM spring become negative, resulting in a CM-V collapse. In this scenario, the voltages of all buses drops nearly identical.

	This case study highlights the effectiveness of the proposed voltage spring metric. It not only indicates the risk of voltage collapse but also points out the specific mode at risk, enabling effective solutions. More specifically, for CM-V, only the total support capability of the devices needs to be improved. In contrast, for DM-V, both device and network influence stability. With stronger device support, the reliance on network strength is weaker, as exemplified in \eqref{Eq_L}.

\section{Discussions and Conclusions}

\subsection{Discussions}

	The study is primarily based on three assumptions: 1) the gird-connected devices are modeled with a unified structure using uniform parameters, 2) the power network is static and 3) F-V coupling is neglected. These assumptions simplify the derivation and aid in understanding the basic concepts of different modal components. Notably, the inertia-damper-spring structure corresponds to the PID control framework, which has been widely validated in industrial applications. 
	Therefore, the analysis based on this structure holds practical and theoretical value in PE-dominated power systems.
	% Specifically, in frequency regulation, inertia represents the derivative term, damping corresponds to the proportional term, and stiffness aligns with the integral term. 

	Notice that, these assumptions are not strictly necessary and can be relaxed. For instance, \cite{zhou2023hetero} and \cite{gao2024vcmf} demonstrate how to approximately decouple systems with heterogeneous devices, where \cite{zhou2023hetero} addresses dynamic networks. Furthermore, \cite{gao2024vcmf} incorporates voltage dynamics into frequency analysis, showing that the coupling can be addressed.  
    Analyzing F/V strength for such complex systems will be our future work.  
	% The underlying principle of these methods is consistent: using eigenvectors to approximate the interactive relationships between heterogeneous devices or between frequency and voltage, then form eigen-subsystems based on these relationships.

\subsection{Conclusions}
	This paper presents a unified framework for F/V strength analysis, illustrating the similarities and differences between frequency and voltage strength. Both F/V responses consist of CM and DM components, namely, CM-F, DM-F, CM-V, and DM-V. The CM-F and CM-V components represent the global frequency or voltage response, and their strength is decided by the collectively active/reactive power support abilities from all devices. In contrast, the DM components capture the spatial differences in the F/V response, and their strength is closely related to the network topology and device locations. 

	In traditional power systems, the strong voltage support provided by SGs makes the CM-V negligible. Additionally, the large DM-F spring constant from strong power network (together with device damping) ensures the rapid suppression of DM-F, leading to CM-F typically dominating the frequency response. As a result, metrics like SCR (closely linked to DM-V spring constant) and total inertia (CM-F inertia) are critical for quantifying F/V strength. However, in PE-dominated power systems, IBRs may fail to maintain voltage source characteristics when they approach their capacity limits, potentially causing CM-V instability. Furthermore, weak interconnections between renewable energy stations and the main grid, combined with insufficient frequency support from IBRs, can lead to non-negligible DM-F responses. 
	Hence, all four types of F/V responses must be considered in the strength analysis for PE-dominated power systems, where both device support and network structure should be taken into account simultaneously.

\normalem
\bibliographystyle{IEEEtran}
\bibliography{references}

\end{document}